\documentclass[
    ,final            
  ]
  {aipproc}

\layoutstyle{6x9}
\usepackage{amssymb,amsmath}

\begin{document}

\title{A model for gamma-ray binaries, based on the effect of pair production feedback in shocked pulsar winds}



\author{E.V. Derishev}{
  address={Institute of Applied physics RAS, 46 Ulyanov st, 603950 Nizhny Novgorod, Russia}
  ,altaddress={Lobachevsky State University of Nizhni Novgorod, 23 Gagarin av, 603950 Nizhny Novgorod, Russia} 
}

\author{F. A. Aharonian}{
  address={School of Cosmic Physics, Dublin Institute for Advanced Studies,
31 Fitzwilliam Place Dublin 2, Ireland}
  ,altaddress={Max Planck Institute f\"ur Kernphysik, Saupfercheckweg 1, D-69117 Heidelberg, Germany} 
}

\begin{abstract}
We analyze the model of gamma-ray binaries, consisting of a
massive star and a pulsar with ultrarelativistic wind. We consider
radiation from energetic particles, accelerated at the pulsar wind
termination shock, and feedback of this radiation on the wind
through production of secondary electron-positron pairs. We show
that the pair feedback limits the Lorentz factor of the pulsar
wind and creates a population of very energetic pairs, whose
radiation may be responsible for the observed gamma-ray signal.
\end{abstract}

\maketitle

\section{Introduction}

Binary systems consisting of a massive star and a compact object
are known to be sources of high-energy gamma-rays. The nature of
compact objects in these systems is not exactly known; there are
models, which either assume they are black holes (e.g.,
\cite{B-R}) or pulsars (e.g., \cite{Dubus}). In the present paper,
we consider gamma-ray binaries with pulsars, where one naturally
expects collision of stellar wind with ultrarelativistic pulsar
wind. In all such models, the pulsar wind is assumed to be the
source of observed gamma-rays -- through inverse Compton
scattering of stellar photons -- either directly, by electrons
moving with extreme bulk Lorentz factor $\sim 10^4 \div 10^5$
(e.g., \cite{Khang}), or from the wind termination shock, where
electrons are accelerated well beyond their initial (bulk) Lorentz
factor $\sim few \times 10^3$ (e.g., \cite{Bark}).

A system, where there is a dense radiation field (here -- stellar
photons) and an ultrarelativistic plasma flow (here --  the pulsar
wind) can launch a runaway-like particle acceleration (converter
acceleration), as been already discussed in application to
gamma-ray bursts and active galactic nuclei (\cite{Der,Stern}). In
gamma-ray binaries the feedback from downstream of the pulsar wind
to its upstream is realized through two-photon pair production.
Below we show that it is possible to construct a model, which can
reproduce main observed properties of gamma-ray binaries without
involving any physical processes apart from the two-photon pair
production feedback itself and consequent cooling of the energetic
pairs through synchrotron and inverse Compton radiation.
Obviously, the physics of gamma-ray binaries is much more complex
than this simple model, but our model demonstrates that the pair
feedback and subsequent converter acceleration has to be included
as an essential, if not the most important, component in any
advanced model for gamma-ray binaries with pulsar winds.

\section{Model description and parameters}

For our model, we assume the following setup (base on the
parameters of LS~5039 and LS~I~+61~303, which can be considered to
be typical representatives of gamma-ray binary population). The
massive star has the luminosity $L_{s} = 3 \times 10^{38}$ erg/s
and the mass-loss rate $\dot{M} = 10^{-6} M_{\odot}$/yr. The mass
outflow is due to stellar wind with velocity $V_{w} \equiv
\beta_{w} c \simeq 0.01 c$. The pulsar orbits the massive star at
the distance $D = 3 \times 10^{12}$ cm and produces a magnetized
ultrarelativistic wind with power $L_{p} = 10^{36}$ erg/s), which
consists of electron-positron pairs and has the Lorentz factor
$\gamma$ (we will show later that $\gamma \sim 10^3$) and
magnetization parameter $\sigma \lesssim 1$.

The pulsar wind termination shock  is located at the distance
$\displaystyle R = \left( \frac{L_{p}}{\beta_{w} \dot{M} c^2}
\right)^{1/2} D$ from the pulsar, where the dynamical pressures of
the two winds are equal. At this distance, the energy density in
the pulsar wind is $w = L_{p}/(4\pi R^2 c)$, what constitutes a
fraction $\epsilon_{w} \simeq 0.5$ of the energy density of the
background thermal photons supplied by the massive star, $w_{b} =
L_{s}/(4\pi D^2 c)$. Thus, neither comptonization of stellar
radiation nor the synchrotron and self-Compton radiation is an a
priori dominant emission mechanism.

The shock-bounded pulsar wind region is filled with stellar
photons and the high-energy radiation produced at the termination
shock, which forms (inside the shock) an isotropic and uniform
photon field. Interacting with each other, these photons
occasionally produce electron-positron pairs, which are picked up
by the pulsar wind, becoming much more energetic, and then lose
energy for radiation on their way to the shock and, to a greater
extent, after crossing the shock and entering the shocked plasma.
The rate of secondary pair production $\dot{N}_{ep}$ does not
depend on the angular distribution of the parent photons as soon
as at least one of the interacting photon fields is isotropic (the
shock-generated photons are always isotropic), and the velocities
of secondary pairs are uncorrelated with the wind velocity. When
pairs with such velocity distribution are picked up by the pulsar
wind, their energy increases on average by the factor $\gamma^2$.

\section{Pair feedback in pulsar winds}

Energetic secondary electrons and positrons readily cool: a
fraction of their energy is radiated before they reach the
termination shock, the rest is radiated in the post-shock region.
The cooling is more efficient in the post-shock region both
because the particles spend more time there and because
synchrotron emission is possible in the shocked plasma in addition
to inverse Compton emission. The synchrotron emission from
secondary pairs supplies additional low-energy target photons for
production of new pairs, forming a feedback loop.

The overall contribution of this feedback to the radiative
efficiency of the shocked pulsar wind can be estimated as
\begin{equation}
\label{effic} \eta_{f} = \frac{\gamma^2 \dot{N}_{ep}
E_{ep}}{L_{p}},
\end{equation}
where $E_{ep}$ is the average total energy of a secondary pair at
the moment of birth. Should the pulsar wind become heavily loaded
with secondary pairs, its Lorentz factor decreases to the value,
which ensures the feedback efficiency $\eta_{f}$ is less than
unity.

Consider two isotropic populations of photons with energies around
$E_1$ and $E_2 = 4 (m_e c^2)^2 / E_1$ ($E_1 > E_2$ for
definiteness), and energy densities $w_1$ and $w_2$. The pair
production rate approximately equals
\begin{equation}
\label{pair_rate} \dot{N}_{ep} = \frac{4 \pi}{3} R^3
\sigma_{\gamma \gamma} \frac{w_1 w_2}{E_1 E_2}\, c =
\frac{\sigma_{\gamma \gamma}}{12 \pi D c}\, \frac{\epsilon_1
\epsilon_2 \epsilon_{w}^{1/2} L_{s}^{5/2}}{(m_e c^2)^2
L_{p}^{1/2}},
\end{equation}
where $\sigma_{\gamma \gamma} \simeq 10^{-25}$~cm$^{-2}$ is the
angle-averaged pair production cross-section, $\epsilon_1$ and
$\epsilon_2$ are the energy densities in two populations of
interacting photons in units of the background stellar radiation
energy density. With increase of $E_1$, the optimal energy of
target photons $E_2$ decreases and, for a given radiation energy
density, the number of target photons (hence opacity) goes up. So,
all other things being equal, the larger is $E_1$ (which
determines the energy of electron-positron pair at birth,
$E_{ep}$), the larger is the feedback efficiency.

Even though the feedback may be controlled by inverse Compton
radiation of primary pairs, the overall radiated power is mainly
due to secondary pairs, which have energies $\sim \gamma^2
E_{ep}/(2\, m_e c^2)$ and upscatter stellar photons inefficiently
because of the Klein-Nishina suppression in the scattering
cross-section. Their main energy-loss channel is synchrotron
radiation in the post-shock region, where secondary pairs form the
cooling distribution and their synchrotron spectrum has the photon
index -3/2. It extends in energy up to
\begin{equation}
\label{spec_end} E_s = \gamma^4 \left( \frac{E_{ep}}{2\, m_e c^2}
\right)^2 \frac{\hbar e B}{m_e c},
\end{equation}
where $B = \left( 2 \epsilon_m \epsilon_w L_s c \right)^{1/2}
/(Dc)$ the magnetic field strength in the post-shock region.

When the pair feedback starts to develop, secondary pairs are born
in interaction of comptonized stellar photons (produced by the
primary electrons) with energy density
\begin{equation}
\label{w1} w_1 \simeq \frac{3 \pi R}{c t}\,
\frac{\gamma}{\gamma_0} w_b = 4 \pi R \sigma_T
\frac{\gamma^2}{\gamma_0} \frac{w_b^2}{m_e c^2},
\end{equation}
where $t$ is the cooling timescale, $\sigma_T$ the Thomson
cross-section, and $\gamma_0$ is the Lorentz factor of the pulsar
wind as it would be in absence of the pair feedback. The target
radiation field is synchrotron radiation of the secondary pairs.
Given the photon index $-3/2$, its energy density at $E_2$ is
\begin{equation}
\label{w2} w_2 = \left( \frac{E_2}{E_s} \right)^{1/2} \eta_{f}\,
\epsilon_w w_b.
\end{equation}
Combining equations Eq.~(\ref{effic}, \ref{pair_rate},
\ref{spec_end}, \ref{w1}, \ref{w2}) we find the self-consistent
Lorentz factor:
\begin{equation}
\label{scgamma} \gamma = \left( 3\pi 2^{1/4}\right) \gamma_0 \,
\frac{T_s^{1/2}}{\sigma_{\gamma \gamma} \sigma_T}\,
\frac{\epsilon_m^{1/4} L_p}{\epsilon_w^{3/4} L_s^{11/4}}\,
D^{3/2}\, m_e c^4 \left( \frac{\hbar^2 e^2}{m_e^2 c^3}
\right)^{1/4}.
\end{equation}
The feedback efficiency enters Eq.~(\ref{scgamma}) inexplicitly,
through the ratio $\gamma/\gamma_0$. With this ratio much less
than unity the feedback efficiency is $\eta_f \sim 1$.

The Lorentz factor, found in this way, depends on the distance to
the massive star, $\gamma \propto D^{3/2}$. Then, the synchrotron
cutoff energy is very sensitive to this distance: $E_s \propto
D^{11}$. If this volatile synchrotron cut-off approaches $m_e
c^2$, then the pair feedback enters a different, more robust mode,
where the synchrotron photons from secondary pairs get absorbed on
themselves, so that the feedback becomes over-efficient and
saturates, keeping $E_s$ somewhat below $m_e c^2$. The
self-consistent Lorentz factor now becomes
\begin{equation}
\label{sy_eq_gamma} \gamma = \frac{2^{1/4} (m_e
c^2)^{1/2}}{T_s^{1/4} (\hbar e)^{1/8}}\, \frac{D^{1/8}}{\left( 2
\epsilon_m \epsilon_w L_s c \right)^{1/16}}\, ,
\end{equation}
where we substituted $E_s$ for $m_e c^2$, keeping in mind that
dependence of the equilibrium Lorentz-factor $\gamma$ on $E_s$ is
very weak. Under typical conditions, Eq.~(\ref{sy_eq_gamma}) gives
$\gamma \simeq 750$.

The pair feedback model predicts that the radiation from pulsar
wind and the shock region contains several spectral components of
different origin. First, it is emission from (slowly cooling)
primary pairs -- inverse Compton from stellar radiation, peaked at
\begin{equation}
E_1^{\rm IC} = 3 \gamma^2 T_s \sim 5\, \mbox{MeV},
\end{equation}
and synchrotron, peaked at $\displaystyle E_1^{\rm sy} = \gamma^2
\frac{\hbar e}{m_e c} \, \frac{\left( 2 \epsilon_m \epsilon_w L_s
c \right)^{1/2}}{Dc} \sim 0.3$~eV, which is overwhelmed by the
radiation of the massive star. Next, there is emission from
secondary pairs, both synchrotron, peaked at $E_2^{\rm sy}  \sim
m_e c^2$, and (less efficient due to the Klein-Nishina effect)
inverse Compton, peaked at $E_2^{\rm IC}  = \gamma^2 m_e c^2 \sim
0.3$~ TeV. The energy of comptonized photons is limited by the
energy of secondary pairs.

\begin{theacknowledgments}

E.V. Derishev acknowledges the support from RFBR grant no.
11-02-00364-a and the program "Non-stationary phenomena in
astrophysical objects" of the Presidium of the Russian Academy of
Science.

\end{theacknowledgments}


\begin{thebibliography}{9}

\bibitem{B-R}
V. Bosch-Ramon, J.~M. Paredes, G.~E. Romero , and M. Rib\'o,
\emph{Astron. Astrophys.} \textbf{459}, L25 (2006).

\bibitem{Dubus}
G. Dubus, \emph{Astron. Astrophys.} \textbf{456},
  801 (2006).

\bibitem{Khang}
D. Khangulyan, F.~A. Aharonian, S.~V. Bogovalov , and M. Rib\'o,
\emph{ApJ} \textbf{752}, L17 (2012).

\bibitem{Bark}
V. Bosch-Ramon, and M.~V. Barkov, \emph{Astron. Astrophys.}
\textbf{535}, A20 (2011).

\bibitem{Der}
E.~V. Derishev, F.~A. Aharonian, V.~V. Kocharovsky, and Vl.~V.
Kocharovsky, \emph{Phys. Rev. D} \textbf{68}, 043003 (2003).

\bibitem{Stern}
B.~E. Stern, and J. Poutanen, \emph{MNRAS} \textbf{383}, 1695
(2008).

\end{thebibliography}
\end{document}